\documentclass[10pt,twocolumn,aps]{revtex4}
\usepackage{graphicx,amsmath,epsfig}
\usepackage{indentfirst,here}

\newcommand{\etal}{{\it et~al. }}
\newcommand{\ie}{{\it i.e.,} }
\newcommand{\eg}{{\it e.g.,} }

\begin{document}

\title{Memory of the Unjamming Transition during Cyclic Tiltings of a Granular Pile}

\author{Deboeuf S.$^1$, Dauchot O.$^2$, Staron L.$^3$, Mangeney A.$^1$, Vilotte J.-P.$^1$}
\affiliation{ $^1$Institut de Physique du Globe de Paris, Paris, FR\\
$^2$Service de Physique de l'{\'E}tat Condens{\'e}, CEA, Saclay, FR \\
$^3$Department of Applied Mathematics and Theoretical Physics, Cambridge, UK}

\begin{abstract}

Discrete numerical simulations are performed to study the evolution of the micro-structure and the response  
of a granular packing during successive loading-unloading cycles, 
consisting of quasi-static rotations in the gravity field between opposite inclination angles.
We show that internal variables, \eg stress and fabric of the pile, 
exhibit hysteresis during these cycles due to the exploration of different metastable configurations. 
Interestingly, the hysteretic behaviour of the pile strongly depends on the maximal inclination of the cycles, 
giving evidence of the irreversible modifications of the pile state 
occurring close to the unjamming transition.
More specifically, we show that for cycles with maximal inclination larger than the repose angle, 
the weak contact network carries the memory of the unjamming transition.
These results demonstrate the relevance of a two-phases description --strong and weak contact networks-- for  
a granular system, as soon as it has approached the unjamming transition.

\end{abstract}
\date{\today}
\maketitle

%----------------------------------------------------------------------
\section{Introduction}
\label{intro}

The discrete nature of granular materials makes their evolutions very complex. 
In response to an external driving force, the macroscopic behaviour of a granular system is determined by the intimate 
interplay between the evolving disordered micro-structure, 
\ie the arrangement of the grains and the geometry of their contact network, 
and the frictional interactions at the contacts, that allow for non trivial force transmissions \cite{Sollichetal97,Roux00,Troadecetal02}. 
Even though significant advances have 
been made in the last decade \cite{JaeNagBeh96,OdaIwashita99,deGennes99a,Rajchenbach00}, 
the underlying question of a physically-based identification of the relevant internal 
variables and of their evolution laws remains open \cite{NematHori99,RadjaiRoux04}. 

A particular issue of recent works on granular media concerns the transition from/to a 
rigid solid-like state to/from a flowing fluid-like state, 
due to its broad interest in geological and industrial processes 
and in particular to its challenge for condensed matter physics.   
This transition , \eg the signature of the development of a yield stress 
or flow threshold in a disordered granular system, or of an infinite relaxation time compared to
the actual experimental time-scale, 
has recently generated a flurry of activity \cite{Knightetal95,OHernetal01,HartleyBehringer03,GdrMidi04}.
Even though dense granular materials are athermal, it has recently been proposed that the transition 
could be seen as a ``unjamming/jamming transition" at zero temperature \cite{LiuNagel98,OHernetal01,OHernetal03a},
motivated by remarkable analogies between granular and glassy systems at both macroscopic 
and microscopic scales \cite{Berthieretal01,Pouetal03,Conetal04,MartyDauchot05}. 

In this picture, the granular system is driven out of equilibrium under quasi-static external driving force, 
introducing 
fluctuations, as the system explores different packing configurations or metastable states, 
that could be related to an ``effective temperature" \cite{Edwards94,Onoetal02}. 
Important non equilibrium effects of structurally disordered granular systems,  
beside the slow relaxation to equilibrium \cite{Deboeufetal03,HartleyBehringer03,UttBeh04}, are :
bistability and hysteresis \cite{MehtaBarker01,Metetal02,DanBeh04} as the result of a memory effect due to  
the internal history dependence of systems out of equilibrium  
and the lack of an unique metastable state \cite{Roux00} ;
and a jerky response to the external driving force  
with impulse-like events (local rearrangements of grains or avalanches), 
as the result of the exploration of local minimal energy configurations \cite{Staronetal02,Kablaetal04}. 
A related important question is whether a diverging length-scale exists on the jammed side 
of the transition \cite{Siletal05}.

The well known observation of different characteristic angles of the stability of 
granular systems results from such memory effects.
When dense granular piles are inclined in the gravity field, 
there exists a maximal angle of stability, or avalanche angle
$\theta_a$ at which the pile starts inevitably to flow, and an angle of repose $\theta_r < \theta_a$, 
defined as the slope angle at which the system comes back to rest.  
In the range $[\theta_r,\theta_a]$, the system exhibits a bistable behaviour where it can be 
in a jammed --rigid-- or an unjammed --flowing-- state. 
In that region, the jammed state of the granular pile is conditionally stable, 
\eg an avalanche can be triggered by perturbations of finite amplitude \cite{Jaegeretal89,DaerrDouady99}, 
evidencing metastability of the pile for slopes in $[\theta_r,\theta_a]$. 
Under rotation in the gravity field, the pile undergoes therefore a subcritical transition. 

In \cite{Staronetal02,Staronetal04,StaronRadjai05}, 
the unjamming transition of a $2D$ cohesionless granular bed slowly tilted towards 
the avalanche angle $\theta_a$ was numerically investigated. 
This transition is characterized by a jerky response
to the smooth rotation of the principal stress direction with respect to the packing, 
as attested by the occurrence of local rearrangements of grains.
An analysis of the local stresses reveals the existence of a significant 
population of overloaded grains, carrying a shear stress ratio larger than 
the critical threshold of the packing at the unjamming transition.
This allows to define a coarse-graining length scale, or ``correlation length", 
which increases with the rotation up to an angle identified to $\theta_r$, 
where it jumps to a length-scale comparable with the thickness of the granular bed. 
This jump can be mapped onto a percolation transition of the overloaded grains, 
and coincides with the onset of a packing dilation, 
indicating coherent  shearing --a macroscopic shear strain across the sample-- prior to the unjamming transition. 
Further insights in the domain $[\theta_r,\theta_a]$ were achieved when 
analysing the highly frictional contacts, \eg the 
critical contacts where the friction is fully mobilized. 
Indeed the critical contacts are at incipient slip and 
likely give rise to local rearrangements of grains.
A ``correlation length", based on the multi-scale 
analysis of the critical contacts, exhibits a power-law type divergence 
with an onset that coincides with $\theta_r$. 
These highly frictional contacts tend to be associated with the weak contact network, as 
defined in \cite{Radjaietal96,Radjaietal98}, suggesting a two-phases system and a second-order phase transition. 
The emergence of long-range ``correlation lengths" within the range $[\theta_r,\theta_a]$, 
both in the stress and the frictional states, is 
consistent with experimental measurements of both the amplitude of 
the local perturbation required to trigger an avalanche and 
the increasing scale of the response to a local perturbation, 
close to the unjamming transition \cite{DaerrDouady99,Deboeufetal03}.   
This suggests that a peculiar regime exists before the unjamming transition, 
namely in the range $[\theta_r,\theta_a]$,
where the accumulation of frictional forces and resulting local rearrangements of grains 
leads to coherent shearing and enhances the structural disorder in the packing \cite{Nasunoetal97}.

Such modifications of the micro-structure of granular systems within the  
range $[\theta_r,\theta_a]$ at the unjamming transition should have a signature on non equilibrium effects, 
such as the system hysteresis during cyclic solicitations. 
In contrast to cyclic shear solicitations \cite{Pouetal03,Toiyaetal04,UttBeh04,Mueggenburg04,AlonsoHerrmann04,Garcia05},   
we focus here --in the continuation of \cite{Staronetal02,Staronetal04,StaronRadjai05}-- 
on quasi-static cyclic rotations of dense cohesionless granular systems under gravity. 
Depending on the amplitude of the cyclic solicitations, the 
granular system can approach the unjamming transition, without overpass it, 
so as to keep a solid-like state and a quasi-static evolution.  
The internal history dependence or hysteresis can therefore 
be investigated depending on how far from equilibrium the system has been driven, 
\ie on the distance from the unjamming transition. 
We report here results of discrete numerical simulations, based on the contact dynamics 
method, of quasi-static cyclic solicitations applied to a $2D$ cohesionless granular bed. 
These results show 
that the hysteresis does depend on the amplitude of the cycles, and exhibit specific memory effects when 
the system approaches the unjamming transition, as a result of the structural evolution. 
In particular, the hysteresis is analysed in terms of both strong and weak contact network
contributions to the global response. 
It is found that the weak contact network carries
most of the signature of the memory effects, 
confirming the relevance of a two-phases description for the unjamming transition, and
extending it to the quasi-static rheology, 
as soon as the granular system has approached the transition.  

The paper is organized as follows. The numerical procedure and details of the numerical simulations 
are presented in section~\ref{setup}. 
The section~\ref{clusters} focuses on critical contacts : 
after discussing previous results during monotonic external loading, we investigate the evolution 
of the critical contacts during relaxation and cyclic solicitations.
In section~\ref{hysteresis} the hysteresis of the granular system is analysed with respect to 
the amplitude of the cycles. 
Then the respective contributions of the weak and strong contact networks to 
the hysteresis of the packing are investigated in section~\ref{bimodal}. 
The paper ends with a discussion of the results in section~\ref{discussion}.

%----------------------------------------------------------------------
\section{Numerical procedures}
\label{setup}

Discrete simulations of granular media
take into account the individual existence of each grain constituting the system. 
The behaviour of the collection of grains is entirely driven by the usual equations of motion 
and the contact laws ruling the interactions between the grains.
The contact dynamics method, applied for this work, deals directly with infinitely stiff contact laws, 
assuming that grains interact through hard-core repulsion and non-smooth Coulomb friction only \cite{JeanandMoreau92}.
This implies that the contact force between two grains is non-zero only if the latter are touching.
Once a contact is created between two grains, the latter cannot get closer, 
so that any normal relative motion is only repulsive : the grains are perfectly rigid and cohesionless.
The Coulomb frictional law consists of an inequality between the tangential and normal forces at contact, 
referred to as $T$ and $N$ respectively : $\mp \mu N\leq T\leq \pm\mu N$,  
where $\mu$ is the microscopic coefficient of friction, 
and sets a threshold for a contact to slip.
Accordingly, this controls the frictional dissipation at the contact scale 
in the case $T=\pm\mu N$ when tangential slip motion is possible 
according to the immediate environment of the contact,
by contrast with the case $\mp \mu N<T<\pm\mu N$ and no slip motion is allowed.
In the present work, we are interested in the quasi-static evolution of the packing
and concentrate on micro-plastic rearrangements only, therefore 
we consider purely dissipative collisions, and set the coefficient of restitution to zero.
%%    Fig 1     %%%%%
\begin{figure}[!ht]
 \begin{minipage}{\linewidth}
  \centerline{ \epsfig{file=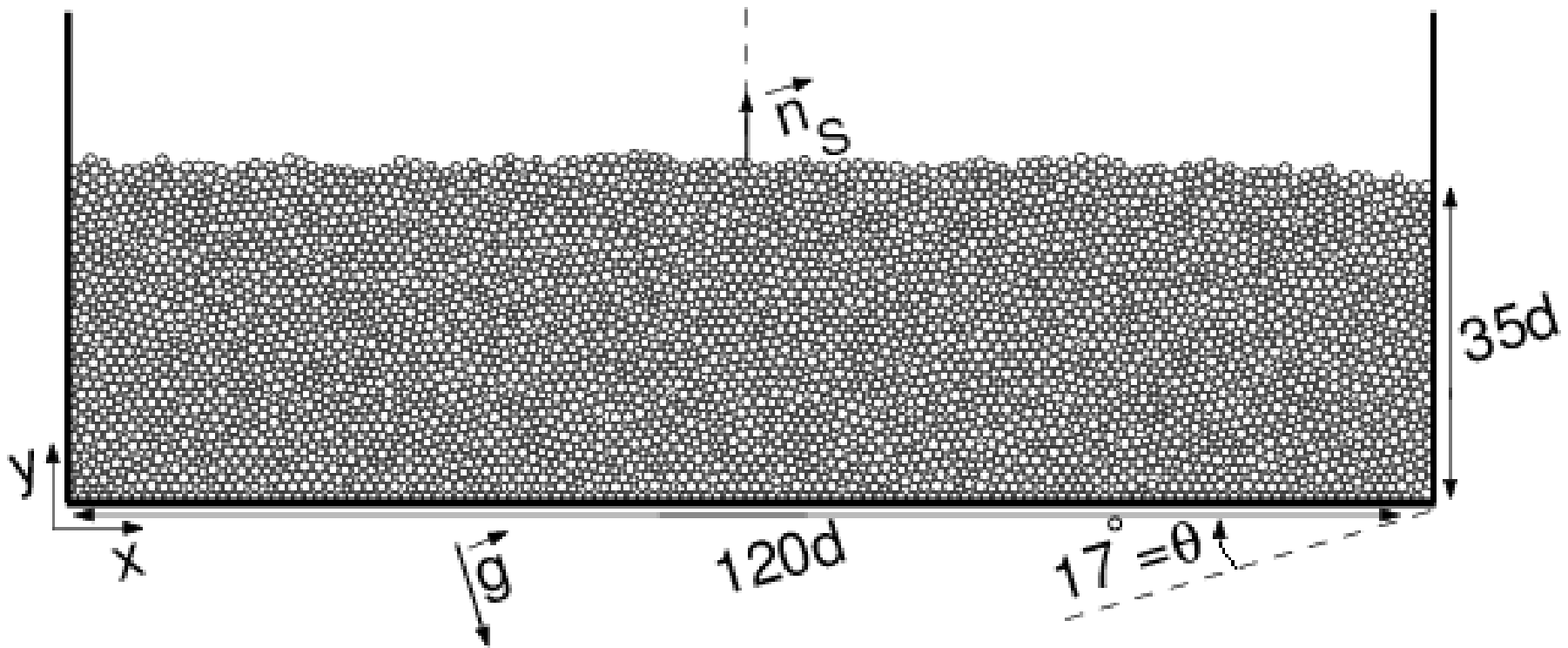,angle=-17,width=\linewidth} }
  \caption[]{A typical simulated granular pile of $4000$ disks whose diameters are polydispersed at $20\%$, 
	inclined at an angle $\theta=17^{\circ}$ in the gravity field $\vec{g}$ with the $xy$-frame 
	linked to the bottom of the box and the unity vector normal to the free surface $\vec{n}_S$.}
  \label{pile}
 \end{minipage}
\end{figure}
%%fin Fig%%%%
%%    Fig 2     %%%%%
\begin{figure}[!ht]
 \begin{minipage}{\linewidth}
  \centerline{ \epsfig{file=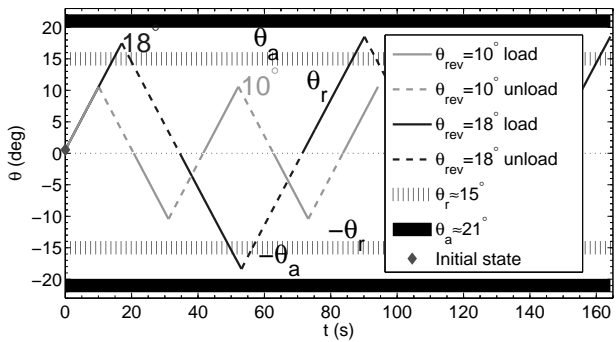,width=\linewidth} }
  \caption[]{Temporal plot of the surface slope $\theta(t)$ of the granular pile 
	during the loading-unloading solicitations for $\theta_{rev}=10^{\circ}$ (in gray) 
	and $\theta_{rev}=18^{\circ}$ (in black). 
	The horizontal lines, with a thickness equal to the size of the fluctuations among $50$ realizations, 
	correspond to the repose
	and avalanche angles $\theta_r$ and $\theta_a$. The initial state of the pile just before the cycles start 
	is represented by a diamond.}
  \label{protocol}
 \end{minipage}  
\end{figure}
%%fin Fig%%%%

The performed simulations are bi-dimensional. 
We consider packings of $4000$ circular grains, randomly deposited in a rectangular box. 
The grains diameters are uniformly distributed in 
the interval $[d_{min};d_{max}]$ with $d_{max}/d_{min}=1.5$. 
The polydispersity, measured as $(d_{max}-d_{min})/(2d)\simeq20\%$ with $d$ the mean disk diameter, 
prevents long-range crystal-like ordering in the packing by enhancing structural disorder. 
Each granular bed shows a nearly flat surface parallel to the bottom of the box, 
with a slope $\theta$ with respect to the horizontal direction ; 
its thickness is $\simeq35d$, and its length is $\simeq120d$. 
The granular packings have an initial solid fraction $C\simeq0.785$ 
and an initial coordination number $Z\simeq3.46$. 
The microscopic coefficient of friction ($\mu=0.5$) is the same between all the grains and 
between the grains and the walls of the box. 
Figure~\ref{pile} shows a typical granular pile inclined in the gravity field $\vec{g}$ 
at an angle $\theta=17^{\circ}$, with the $xy$-frame linked to the bottom of the box and
the unity vector normal to the free surface $\vec{n}_S$. 
In practice, instead of rotating the granular pile in the constant gravity field,  
the direction of $\vec{g}$ with respect to the pile is rotated to simulate tilting.  

The cyclic solicitations consist of rotations at a constant rate 
($\pm 0.001^{\circ}$ per time step), the rotation being positive when clockwise.
A positive rotation is first applied to newly generated granular piles 
so that the slope of the free surface increases from $\theta=0^{\circ}$ 
to a maximal inclination angle $\theta_{rev}$, corresponding to a loading stage. 
After the inclination $\theta_{rev}$ is reached, the direction of rotation is reversed, 
and the slope decreases back to $0^{\circ}$, corresponding to an unloading stage. 
Rotation is maintained so that the slope of the pile evolves from $0$ 
towards the opposite maximal inclination angle $-\theta_{rev}$. 
The direction of rotation is again reversed, and the cyclic solicitation is resumed. 
Two successive cycles are performed.
This solicitation is applied on $50$ granular piles differing 
in the initial disordered micro-structure, \ie the grains arrangement and the contact network geometry, 
and showing characteristic angles $\theta_r\simeq15^{\circ}$ and $\theta_a\simeq21^{\circ}$.
In the following, averages are taken over all $50$ simulations.  
The granular beds and the data describing the first loading stage, 
are the actual data of Staron \etal \cite{Staronetal04}.

We study the influence of the cycle amplitude and of the distance from the unjamming transition  
on the evolution of the pile state by 
considering two values of the maximal inclination angle $\theta_{rev}$, 
namely by applying small and large cycles.
The smallest value of $\theta_{rev}$ is chosen so that the small cycle keeps 
the pile out of the coherent shear regime identified prior to the unjamming transition : 
$\theta_{rev}=10^{\circ}<\theta_r$.  
On the contrary, the largest value of $\theta_{rev}$ is chosen in the range of slopes $[\theta_r;\theta_a]$, 
but below the minimal avalanche angle observed among the $50$ realizations 
to prevent the pile from avalanching :  $\theta_{rev}=18^{\circ}\in[\theta_r;\theta_a]$. 
These two kinds of cycles are represented in Figure~\ref{protocol}, 
where the slope of the pile $\theta$ is plotted against the time $t$ for 
both small (in gray) and large (in black) cycles.  
Continuous lines represent loading stages, and dashed lines represent unloading ones.
The horizontal lines indicate the repose and avalanche angles $\theta_r$ and 
$\theta_a$, with a thickness equal to the size of the fluctuations.
A diamond indicates the initial state of the granular pile just before the cycles start, that will be reported 
on all the next figures.

%----------------------------------------------------------------------
\section{Critical contacts clustering}
\label{clusters}

%%    Fig 3       %%%%%%%
\begin{figure}[!ht]
 \begin{minipage}{\linewidth}
  \centerline{ \epsfig{file=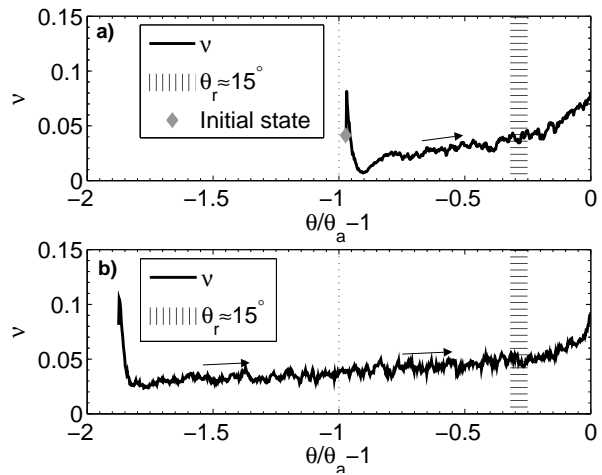,width=\linewidth}}
 \end{minipage}
 \caption[]{Density of critical contacts $\nu$ averaged over $50$ simulations (a) during the
        first loading starting from $\theta\simeq0^{\circ}$ and (b)
        during  one large cycle, starting from a pile inclined at $-\theta_{rev}= -18^\circ$,
        when varying the slope inclination towards the avalanche angle.}
 \label{nu}
\end{figure}
%%fin Fig%%%%%%%%

As discussed in the case of slow monotonic external loading of a granular slope 
in \cite{Staronetal02,Staronetal04,StaronRadjai05},  
the coherent shear regime identified prior to the unjamming transition corresponds to  
the structuration in the packing of highly frictional contacts, 
\eg contacts where forces have reached the Coulomb frictional threshold, 
\ie $T=\pm\mu N$, referred to as critical contacts. 
An interesting issue would be the evolution of the critical contacts 
during cyclic solicitations.

We discuss first the robustness of previous results on critical contacts \cite{Staronetal02,StaronRadjai05} against the pile preparation. 
Figure~\ref{nu} displays the evolution of the density of critical contacts $\nu$, 
defined as the proportion of critical contacts to the total number of contacts 
in a granular packing, averaged over $50$ piles, when varying the pile inclination 
$\theta$ towards $\theta_a$, from $0^{\circ}$ $(a)$ and from $-\theta_{rev}= -18^\circ$ $(b)$. 
At the beginning of the solicitation, both evolutions of $\nu$ show a sudden drop to a minimum value
indicating the instantaneous loss of critical contacts. 
After this first stage and in spite of the different initial conditions, 
the behaviour of the critical contacts when approaching the unjamming transition is approximately the same : 
$\nu$ increases up to a maximal value $\nu_c\simeq8\%$ \cite{Staronetal02}. 

The loss of critical contacts occurring at the transition 
from loading to unloading in Figure~\ref{nu}$b$, also observed in \cite{AlonsoHerrmann04}, 
indicates that the  {\it averaged} frictional state 
of the packing related to past solicitations is immediately modified. 
It is related to a systematic grains rearrangement, 
during which critical contacts leave the Coulomb frictional threshold.  
Note that the amount of  critical contacts lost at the transition is not larger 
than the typical fluctuations around the averaged critical contacts density $\nu$ during the cycle. 
In fact, the transition from loading to unloading imposed at $\theta_{rev}$ synchronises the grains rearrangements 
for the various realizations, hence a larger response on average than in the other parts of the cycle. 
The effect of the reversal on a single realization remains an open issue, 
which is beyond the present averaged analysis.  
 
%%    Fig 4       %%%%%%%
\begin{figure}[!ht]
 \begin{minipage}{\linewidth}
  \centerline{ \epsfig{file=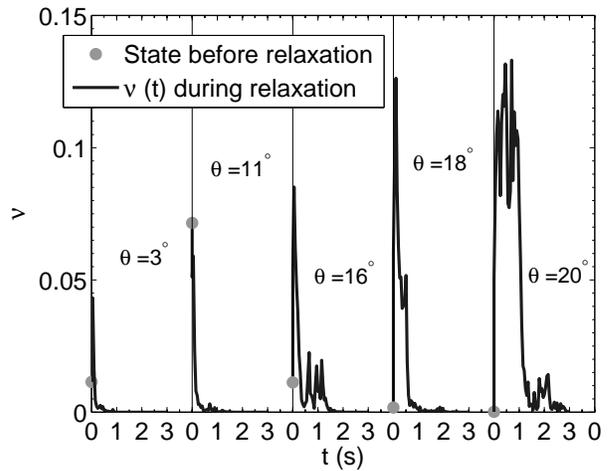,width=\linewidth} }
 \end{minipage}  
 \caption[]{Relaxation of the density of critical contacts $\nu$ as a function of time in 
	$5$ granular piles inclined at different angles ($\theta=3^\circ,11^\circ,16^\circ,18^\circ,20^\circ$)
	after the rotation was stopped at $t=0$. The state of the pile just before the relaxation starts is 
	represented by a circle.}
 \label{nuzero}
\end{figure}
%%fin Fig%%%%%%%%

We now observe the relaxation to a static packing in terms of critical contacts 
by suddenly stopping the rotation. 
Figure~\ref{nuzero} shows the evolution of $\nu$ as a function of time after the rotation was stopped at $t=0$
in $5$ granular piles inclined at different angles ($\theta=3^\circ,11^\circ,16^\circ,18^\circ,20^\circ$). 
The state of the pile before the relaxation starts is represented by a circle. 
Whatever the density of critical contacts and 
the inclination when the packing starts to relax, $\nu$ vanishes after a complex transient dynamics. 
The time needed for critical contacts to relax is larger for slopes close to $\theta_a$ : this confirms  
the complex relaxation towards equilibrium of a granular packing \cite{Deboeufetal03,HartleyBehringer03,UttBeh04}, and 
this is consistent with observations of long-range correlations close to the unjamming transition \cite{Jaegeretal89,DaerrDouady99,Staronetal02,Deboeufetal03,Staronetal04}. 
The relaxation of the critical contacts for rigorous static conditions shows that 
the critical state is a transient state. 
This can be understood by the fact that the contacts can not sustain 
the sliding condition once the loading is stopped. 
In this sense, the density of critical contacts does not  
characterize the pile plastic state, 
because $\nu$ is  a dynamic response function to the actual loading of the pile.
Also other variables  could be more representative of the plastic state of the pile, \eg  the averaged cumulative slip dislocation at critical contacts. 
However critical contacts remain active during quasi-static loading, and it was shown that 
the evolution of $\nu$ for one realization \cite{Staronetal02} 
becomes more and more intermittent, exhibiting successive frictional demobilization and  remobilization periods, when the rotation rate decreases. 

%%    Fig 5       %%%%%%%
\begin{figure}[!ht]
 \begin{minipage}{\linewidth}
  \centerline{ \epsfig{file=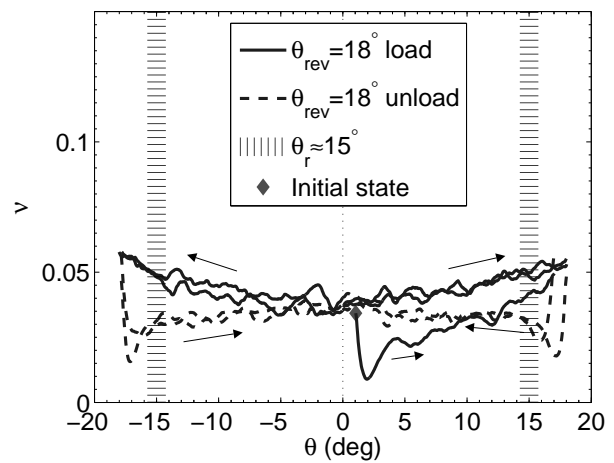,width=\linewidth} }
 \end{minipage}  
 \caption[]{Density of critical contacts $\nu$ averaged over $50$ simulations,  slide-averaged over $1^{\circ}$-intervals 
	as a function of the slope angle $\theta$ during the successive cyclic solicitations for $\theta_{rev}=18^{\circ}$.}
 \label{nucycle}
\end{figure}
%%fin Fig%%%%%%%%
%%    Fig 6       %%%%%%%
\begin{figure}[!ht]
 \begin{minipage}{\linewidth}

  \centerline{ \epsfig{file=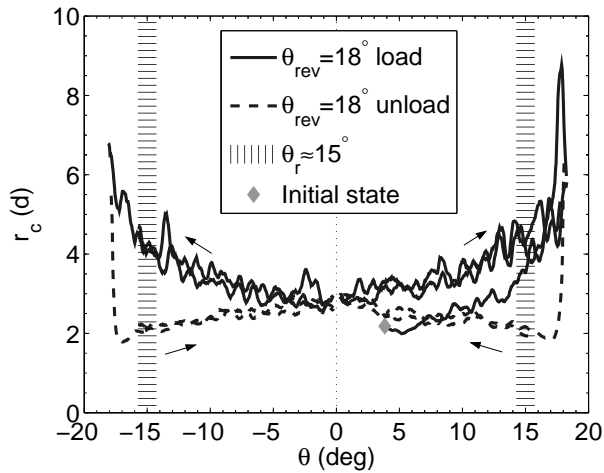,width=\linewidth} }
 \end{minipage}  
 \caption[]{Size of the correlated 
	clusters of critical contacts $r_c$, slide-averaged over $1^{\circ}$-intervals as a function of the slope 
	angle $\theta$ during the successive cyclic solicitations for $\theta_{rev}=18^{\circ}$.}
 \label{corr}
\end{figure}
%%fin Fig%%%%%%%%

 Figure~\ref{nucycle} shows the evolution of  $\nu$ as a function of the pile slope
during loading and unloading process for $\theta_{rev}=18^{\circ}$. 
The density of critical contacts at an angle $\theta$ is smaller during unloading than 
during loading, as attested by the hysteretic loops. 
Figure~\ref{corr} displays the evolution as a function of  $\theta$ 
during the large cycles, of the ``correlation length" $r_c$, defined as the mean size of clusters
such that locally $\nu = \nu_c$. 
During each loading stage, $r_c$ increases sharply above $\theta_r$ and is expected to 
reach the size of the pile at the unjamming transition, as shown 
in the case of monotonic external loading \cite{Staronetal02}. 
At the transition from loading to unloading, a signature  of the loss of critical contacts 
is visible :  
$r_c$ immediately reduces to a few grain diameters. 
There is no more evidence of correlation within the granular pile in terms of critical contacts 
upon reversal of the solicitation. 
Note that after the first loading, $\nu$ as well as $r_c$ shows the same cyclic evolution, indicating 
the same behaviour of critical contacts during the two successive cycles, 
especially when approaching the unjamming transition. 
 The reduction of frictional mobilization after the reversal, 
induces a strong hysteresis in both the evolution of $\nu$ and $r_c$ during loading and unloading cycles. 
Is this hysteresis apparent on other global characteristics of the granular pile?

The critical contacts play a significant role when approaching the unjamming transition, 
since they likely give rise to micro-plastic events. 
Accordingly, one may wonder whether the response of the pile is aed by the hysteretic evolution of 
critical contacts observed in the range of slopes $[\theta_r;\theta_a]$.
Also one would like to know whether the possible effect of the observed hysteresis is localized 
in the domain $[\theta_r;\theta_a]$ or conditions the response of the pile along the full cycle of solicitation.
In the following, we investigate the hysteretic response of the pile to cyclic solicitations and the effect 
on this response of having approached the unjamming transition.
To do so, we consider two different amplitudes of loading-unloading cycles, namely two values of 
the reversal angle $\theta_{rev}$, 
and we compare the response of the granular pile in terms of stress, strain and evolution of the contact network
for $\theta_{rev}=10^{\circ}$ and $\theta_{rev}=18^{\circ}$.

%----------------------------------------------------------------------
\section{Hysteretic phenomena during loading and unloading solicitations}
\label{hysteresis}

During cyclic tiltings,
the induced stresses as well as the resulting deformations of the granular pile are investigated. 
We start with the comparison of the global stress state during loading and unloading stages.

\subsection{Stress state of the granular pile} 

 The maximal value of the shear to confinement ratio, the so-called inertial number in \cite{GdrMidi04} is defined as $I_{max}=\gamma_{max} d{\sqrt\frac{\rho}{P}}$, 
with $\gamma_{max}$, the maximal shear strain rate evaluated during coherent shearing       
in the range of slopes $[\theta_r;\theta_a]$,  
$\rho$, the volumetric mass of the grains and 
$P$, the typical pressure induced by gravity. 
In the present case, $I_{max}\simeq10^{-5}$ shows that the pile evolution is actually quasi-static.  
As a result, static stresses transmitted via contacts are much larger than kinematic ones 
due to momentum transport during collisions, 
so that the stress tensor of the granular pile $\boldsymbol{\sigma}$ in the $xy$-frame
is evaluated  as follows from \cite{KruytRothenburg96} :
\begin{equation} 
\sigma_{ij} = \frac{1}{V}\sum_{\alpha=1}^{n^c}{ f^{\alpha}_i l^{\alpha}_j},
\label{eqstress}
\end{equation}
where $i$ and $j$ denote the space dimensions and $\alpha$ the contact. 
$\vec{f}$ is the contact force, $\vec{l}$ is the vector normal to the contact surface,   
and $n^c$ is the total number of contacts  in the representative element volume $V$. 
The tangential and the normal stresses $\sigma_T$ and $\sigma_N$ 
along the direction of the free surface are calculated :
\begin{eqnarray}
\sigma_N & = & \parallel(\boldsymbol{\sigma}\vec{n}_S)\vec{n}_S\parallel,  \\ 
\sigma_T & = & \parallel\boldsymbol{\sigma}\vec{n}_S-\sigma_N\vec{n}_S\parallel,
\end{eqnarray} 
where $\vec{n}_S$ is the unity vector normal to the free surface.  
The direction of the eigenvector of $\boldsymbol{\sigma}$ corresponding to the largest eigenvalue is 
the principal stress direction, whose angle with the direction $\vec{n}_S$  is $\Psi$. 
The principal stress direction corresponds to the direction where stress is purely compressive.

%%    Fig 7     %%%%%
\begin{figure}[!ht]
 \begin{minipage}{\linewidth}
  \centerline{ \epsfig{file=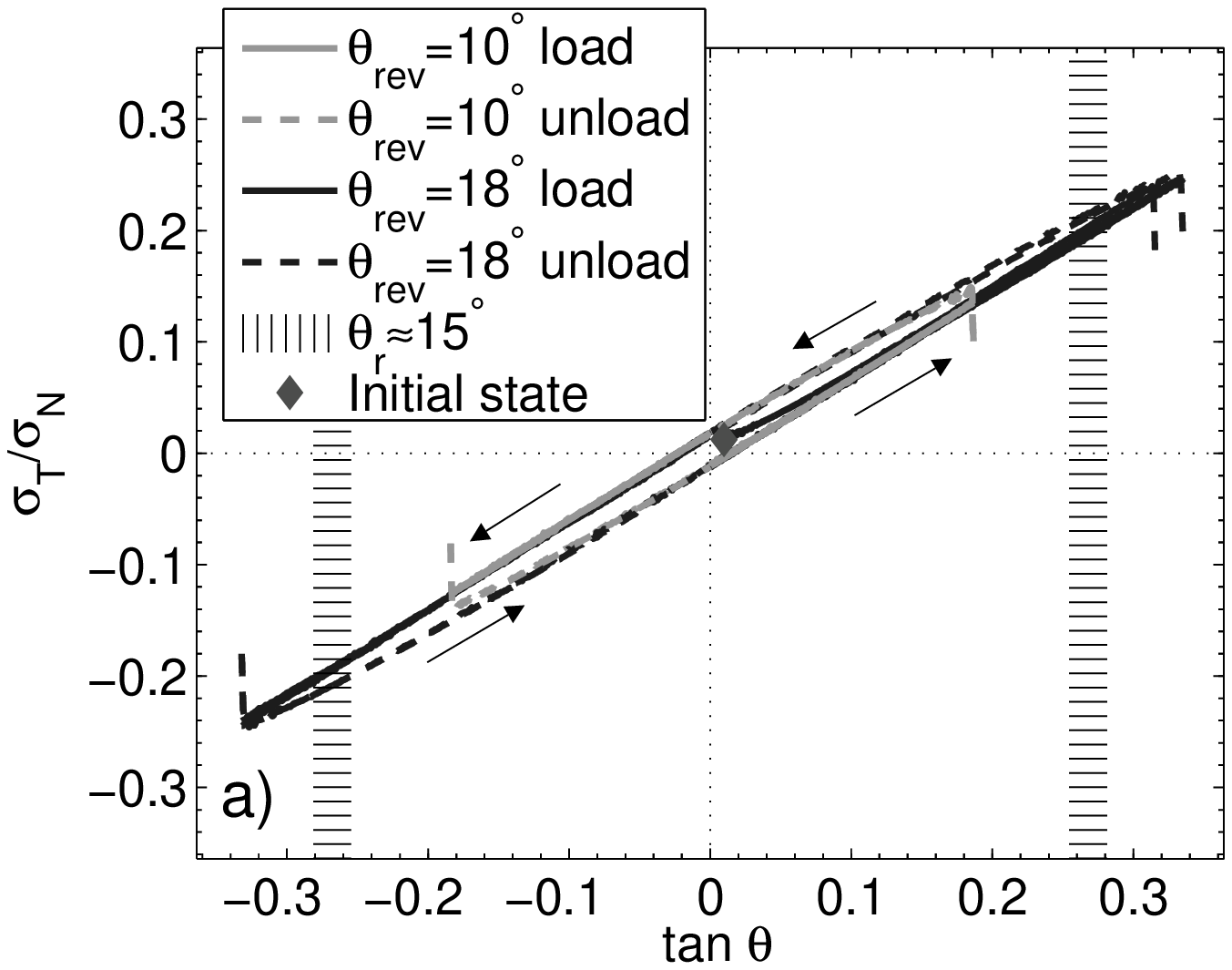,width=\linewidth} }
 \end{minipage}
 \\  
\begin{minipage}{\linewidth} 
  \centerline{ \epsfig{file=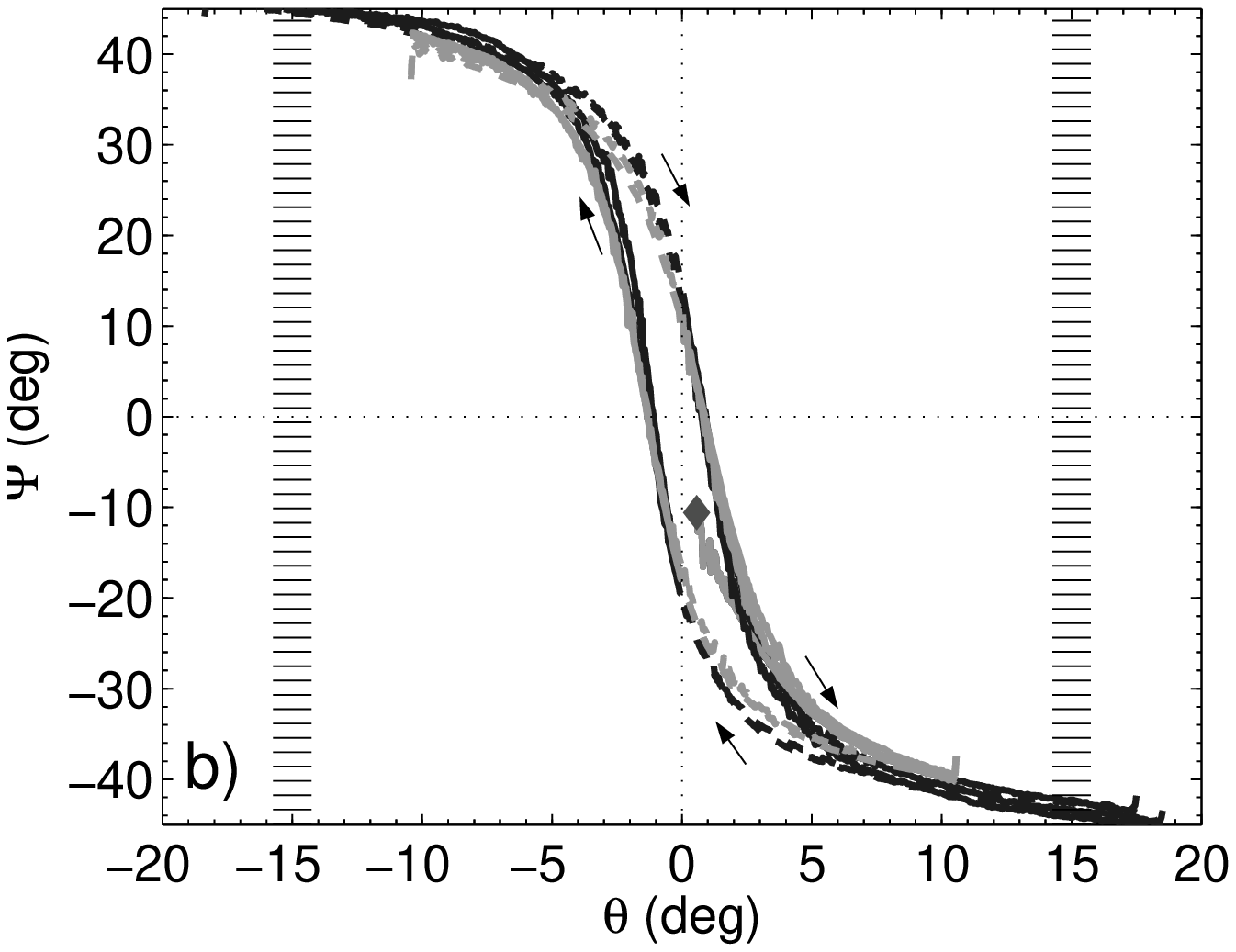,width=\linewidth} }
 \end{minipage}  
 \caption[]{Shear stress ratio $\sigma_T/\sigma_N$ ($a$)  and  
	principal stress direction $\Psi$ ($b$) as a function of the pile slope $\theta$
	during cyclic rotations 
	for small ($\theta_{rev}=10^{\circ}$) and large ($\theta_{rev}=18^{\circ}$) amplitudes.
	The vertical lines correspond to the value of $-\theta_r$ and $\theta_r$.
	The initial state is represented by a diamond.}
 \label{stress}
\end{figure}
%%fin Fig%%%%

Figure~\ref{stress}$a$ displays the shear stress ratio $\sigma_T/\sigma_N$ and Figure~\ref{stress}$b$ 
the evolution of the principal stress direction $\Psi$ 
during small ($\theta_{rev}=10^{\circ}$) and large ($\theta_{rev}=18^{\circ}$) loading-unloading cycles. 
Both $\sigma_T/\sigma_N$ and $\Psi$ exhibit hysteretic evolutions related to irreversible modifications
of the force transmissions at contacts, but the hysteresis loops are of rather small amplitude and
independent of $\theta_{rev}$.

During the first loading, $\sigma_T/\sigma_N\simeq a \tan(\theta)$ 
in agreement with the predictions of continuous media mechanics 
for an infinite slope inclined in the gravity field at static equilibrium,  
except that $a$ is not exactly equal to one : $a=0.8$. 
It can be shown that this difference is to be 
attributed to finite size and walls effects \cite{Staronetal04}. 
During the following cycles, $\sigma_T/\sigma_N\simeq a \tan(\theta) + \epsilon(\sigma_T/\sigma_N)_{res}$,
where $\epsilon= -1$ (resp. $\epsilon= +1$) during loading (resp. unloading) and $a=0.8$.
A residual stress ratio $(\sigma_T/\sigma_N)_{res}$ of the
order of $5\%$ of the typical shear stress ratio is recorded for $\theta\simeq0^{\circ}$. 
The principal stress direction $\Psi$ tends to $\pm45^{\circ}$ for large inclinations,
exhibiting a rather smooth evolution.  
For hydrostatic equilibrium, \ie equal normal stresses : $\sigma_{xx}=\sigma_{yy}$, 
it would be expected that $\Psi=-45^{\circ}$ (resp. $\Psi=+45^{\circ}$) for $\theta>0$ (resp. $\theta<0$).

The stress state evolution of the granular pile is plotted on the figures~\ref{stress} for 
the two successive cycles : the plots perfectly collapse after the first loading.
In the following, the two successive cycles are reported on all the figures.

 Altogether both the shear stress ratio $\sigma_T/\sigma_N$ and 
the principal stress direction $\Psi$ are controled by the inclination $\theta$ (and its temporal derivative, as evidenced by the weak hysteresis), whatever the solicitations path
and the history of past cycles. 

\subsection{Deformation of the granular pile }  
\begin{figure}[!ht]
 \begin{minipage}[b]{\linewidth}
  \centerline{ \epsfig{file=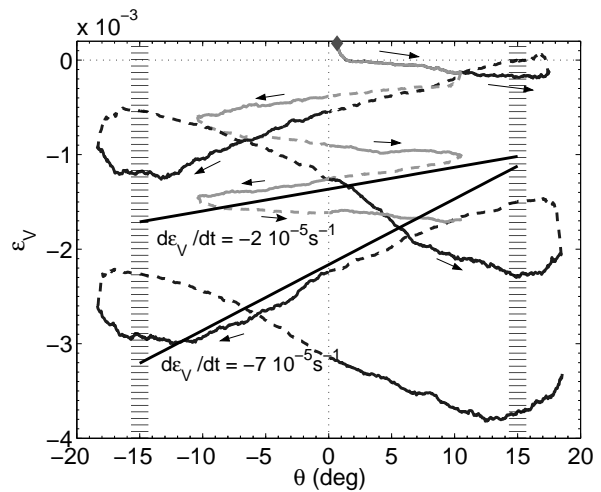,width=\linewidth} }
  \caption[]{ Volumetric strain $\epsilon_V$ as a function of the 
	pile slope during small and large cycles (same symbols as figure~\ref{stress}).
	The two lines are indications of the compaction rate $d\epsilon_V/d t$  
	for $\theta_{rev}=10^{\circ}$ and $\theta_{rev}=18^{\circ}$.}
  \label{epsv}
 \end{minipage}
\end{figure}

The exploration of different metastable configurations results 
in the apparition of critical contacts in the packing, that induces local rearrangements of grains 
(see section~\ref{clusters} and \cite{Staronetal02}). 
The effect of these rearrangements at the scale of the pile is apparent when plotting the variations of the 
volumetric strain $\epsilon_V=(V-V_0)/V_0$, where $V$ is the volume of the pile and 
$V_0$ its initial volume, as a function of the slope of the pile during 
successive small and large cycles (Figure~\ref{epsv}). 
 At the first sight, we notice that, even if the initial solid fraction $C$ and the stress state evolution of the pile are  the same for the two cyclic amplitudes (Fig.~\ref{stress}), the volumetric strain evolution does depend strongly on $\theta_{rev}$. 

%%    Fig 8     %% 
%%fin Fig%%

Irrespective of the value of $\theta_{rev}$, 
cycles produce an overall densification of the pile of the order of $10^{-3}$, 
corresponding to a slight increase of the solid fraction $C$ from $0.785$ to $0.79$.
Obviously, the small number of cycles experienced in the present work does not allow 
to observe a saturation of the volume and 
eventually a critical state \cite{RadjaiRoux04,Porionetal93}. 
It is known that in similar cyclic solicitations \cite{AlonsoHerrmann04}, 
as well as in other experimental set-ups \cite{Knightetal95,Nicolasetal00},
even after a large number of cycles ($10^4$ cycles), the volume still slowly decreases. 

For slopes $\theta\lesssim\theta_r$, the granular pile is contracting :
$\epsilon_V$ decreases.
On the contrary, for slopes in the range
$[\theta_r;\theta_a]$, the rearrangements cause the granular pile to  
dilate : $\epsilon_V$ increases. 
As a result, for small cycles ($\theta_{rev}=10^\circ$) 
the  compaction of the granular pile remains monotonous (Fig.~\ref{epsv}), 
whereas for large cycles ($\theta_{rev}=18^\circ$), the granular pile exhibits
compaction and dilation stages (Fig.~\ref{epsv}). 
This latter behaviour is typical when shearing dense 
granular media, because shear deformation is possible only if grains unjamm, 
confirming the onset of a coherent shearing above $\theta_r$. 

Interestingly, the overall behaviour of the pile over the complete solicitation 
is modified in the case of large cycles : 
the instantaneous compaction is twice more efficient for large cycles ($d\epsilon_V/d t\simeq-7~10^{-5}s^{-1}$) 
than for small ones ($d\epsilon_V/d t\simeq-2~10^{-5}s^{-1}$). 
As a result and despite the dilation stage occurring when $\theta\in[\theta_r;\theta_a]$,   
the overall densification is larger for large cycles  than for small ones for the same number of cycles,   
showing that densification depends on the solicitation path \cite{Joer98}.   
This leads us to speculate that grains are much more allowed to rearrange  
when the pile has previously evolved in the coherent shear regime identified prior to the unjamming transition. 

%%    Figure     %% 
\begin{figure}[!ht]
 \begin{minipage}[b]{\linewidth} 
  \centerline{ \epsfig{file=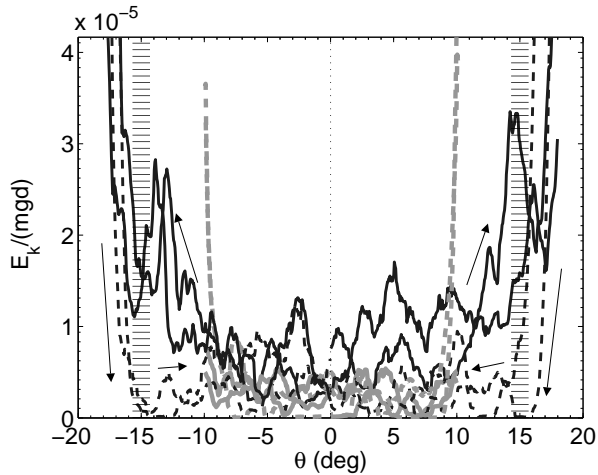,width=\linewidth} }
  \caption[]{Kinetic energy $E_k$ averaged over all the grains and slide-averaged over 
	$1^{\circ}$-intervals, normalised by the typical potential energy of a grain of mass $m$ for a height $d$, as a function of $\theta$ for $\theta_{rev}=10^{\circ}$ and $\theta_{rev}=18^{\circ}$ 
	(same symbols as figure \ref{stress}). }
  \label{ec}
 \end{minipage}
\end{figure}
%%fin Figure%%

 This largest susceptibility for grains to rearrange
in the case of large cycles ($\theta_{rev}=18^{\circ}$) is confirmed by monitoring the 
averaged kinetic energy of the grains  $E_k$, 
represented as a function of the pile slope in Figure~\ref{ec} after slide-average over $1^{\circ}$-intervals. 
The kinetic energy $E_k$ involved in the local rearrangements, also occurring at small inclinations $\theta$,  
is much higher when the granular pile has been loaded up above $\theta_r$ during its history. 
This evolution underlines the two different states and evolutions of the pile for the two cyclic amplitudes.  

Exploring the domain $[\theta_r;\theta_a]$,  
allows for stronger rearrangements and larger volumetric strains 
during the full solicitation despite no apparent signature on $\sigma_T/\sigma_N$. 
As a result, the volumetric deformation of the granular pile shows a 
strong dependence on the amplitude of the cycles. 

\subsection{Evolution of the contact network}
We now investigate the effect of  
the deformations on the evolution of the micro-structure, \ie the geometry of the contact network, 
 that are candidates at relevant internal variables \cite{NematHori99,RadjaiRoux04,ThorntonBarnes86,BathurstRothenburg88,Nemat00,Madadietal04,Alonsoetal05}. 

\subsubsection{Coordinancy of the grains }
%%    Fig 10     %%%%%
\begin{figure}[!ht]
 \begin{minipage}{\linewidth}
  \centerline{ \epsfig{file=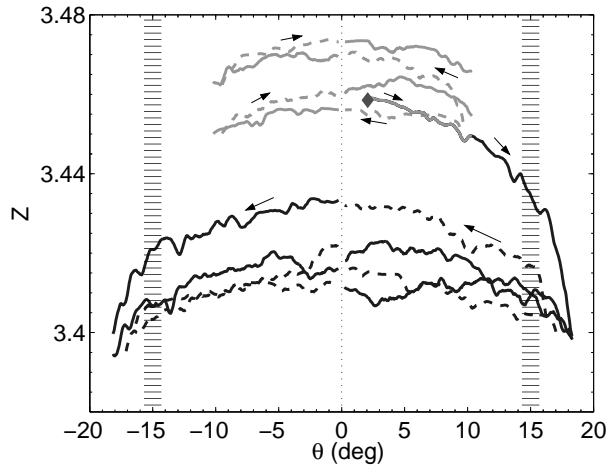,width=\linewidth} }
  \caption[]{Coordinancy of the grains $Z$ slide-averaged over $1^{\circ}$-intervals
	as a function of the slope of the pile for successive small and large cycles 
	(same symbols as figure~\ref{stress}).}
  \label{coord}
 \end{minipage}
\end{figure}
%%fin Fig%%%%

The coordinancy $Z$, \ie the mean number of contacts per grain,  
slide-averaged over $1^{\circ}$-intervals 
is plotted in Figure~\ref{coord} as a function of the pile slope 
during the cyclic solicitations.
For both values of the amplitude $\theta_{rev}$, the coordinancy  
$Z$ varies less than $1\%$. 
Nevertheless, it exhibits a weak hysteresis according to the sense of rotation :
$Z$ tends to decrease during loading, but to increase during unloading.

Small and large cycles are identified by two different states :
the value of the coordinancy depends on $\theta_{rev}$.
During small cycles ($\theta_{rev}=10^{\circ}$), $Z$ remains approximately at its initial value,
by contrast with large cycles ($\theta_{rev}=18^{\circ}$), for which a variation of the 
coordinancy occurs during the first exploration of the domain $[\theta_r;\theta_a]$ 
and remains afterwards. 

Smallest coordinancy observed for $\theta_{rev}=18^{\circ}$ corresponds to densest granular piles, 
contrary to intuitive observations \cite{Siletal02}.
Such an anti-correlation between the evolutions of the coordinancy and the volume under an external driving field
could be related to the dependence of the packing equilibrium on its number of contacts.
An ``isostatic" packing has the minimal number of contacts required to remain at metastable equilibrium,
by contrast with a ``hyperstatic" one, that could loose some contacts and not its equilibrium. 
 Yet for an initially static grain to rearrange, its contacts have to cooperate,  
that is easier for a smaller number of contacts.  
In the hyperstatic packing, a significant cooperation of contacts is needed for a grain to rearrange, 
by contrast with the isostatic packing, in which a slight change in the force transmission at contacts will enable  
a local rearrangement of grains. 
In the case of large cycles, the significant decrease of $Z$ observed during the first loading  
makes the packing become less hyperstatic :
grains rearrangements occur easily, enhancing the deformation of the pile. 

\subsubsection{Fabric of the pile}
%%    Fig 11     %%%%%
\begin{figure}[!ht]
 \begin{minipage}{\linewidth}
  \centerline{ \epsfig{file=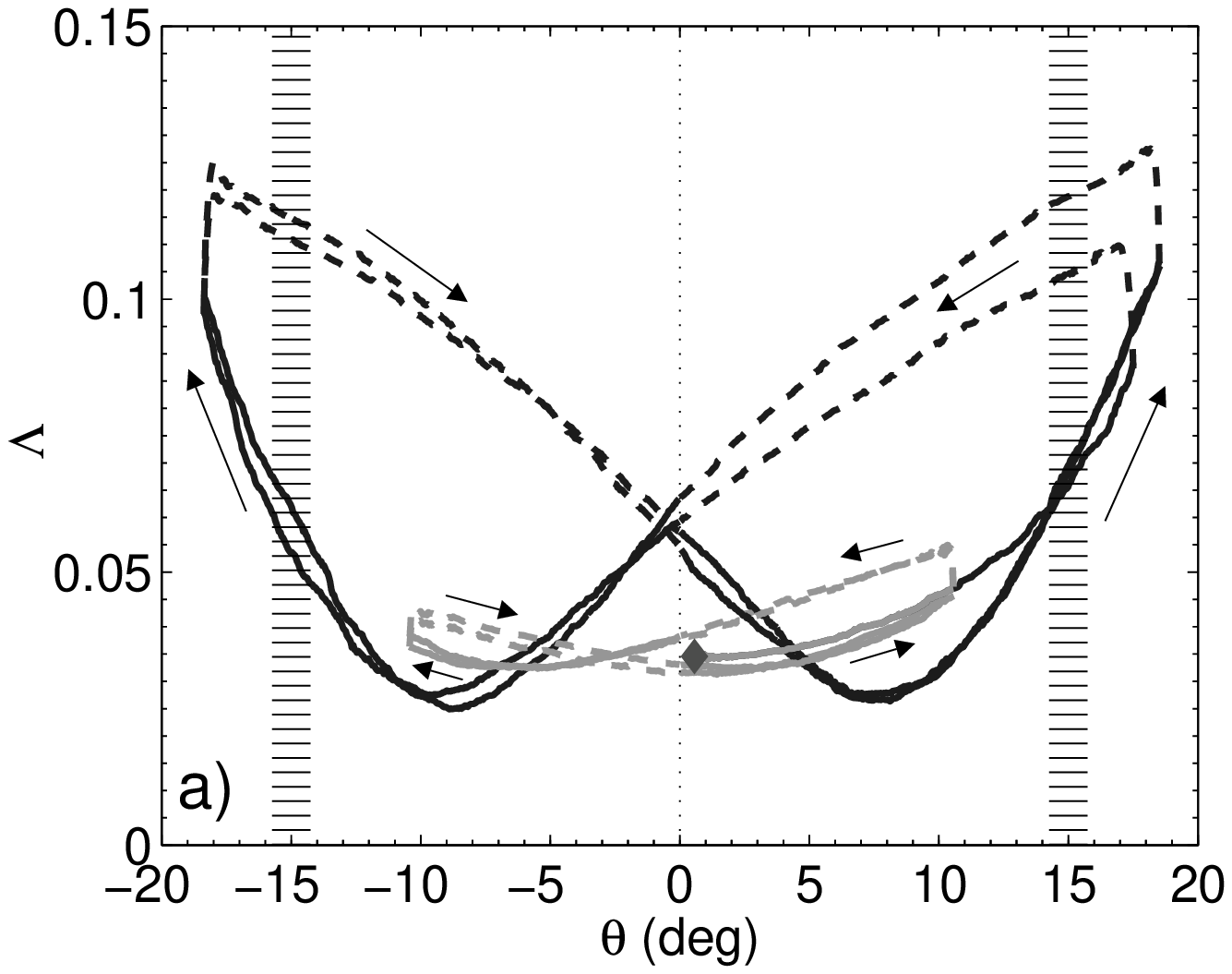,width=\linewidth} }
 \end{minipage}
 \\  
 \begin{minipage}{\linewidth} 
  \centerline{ \epsfig{file=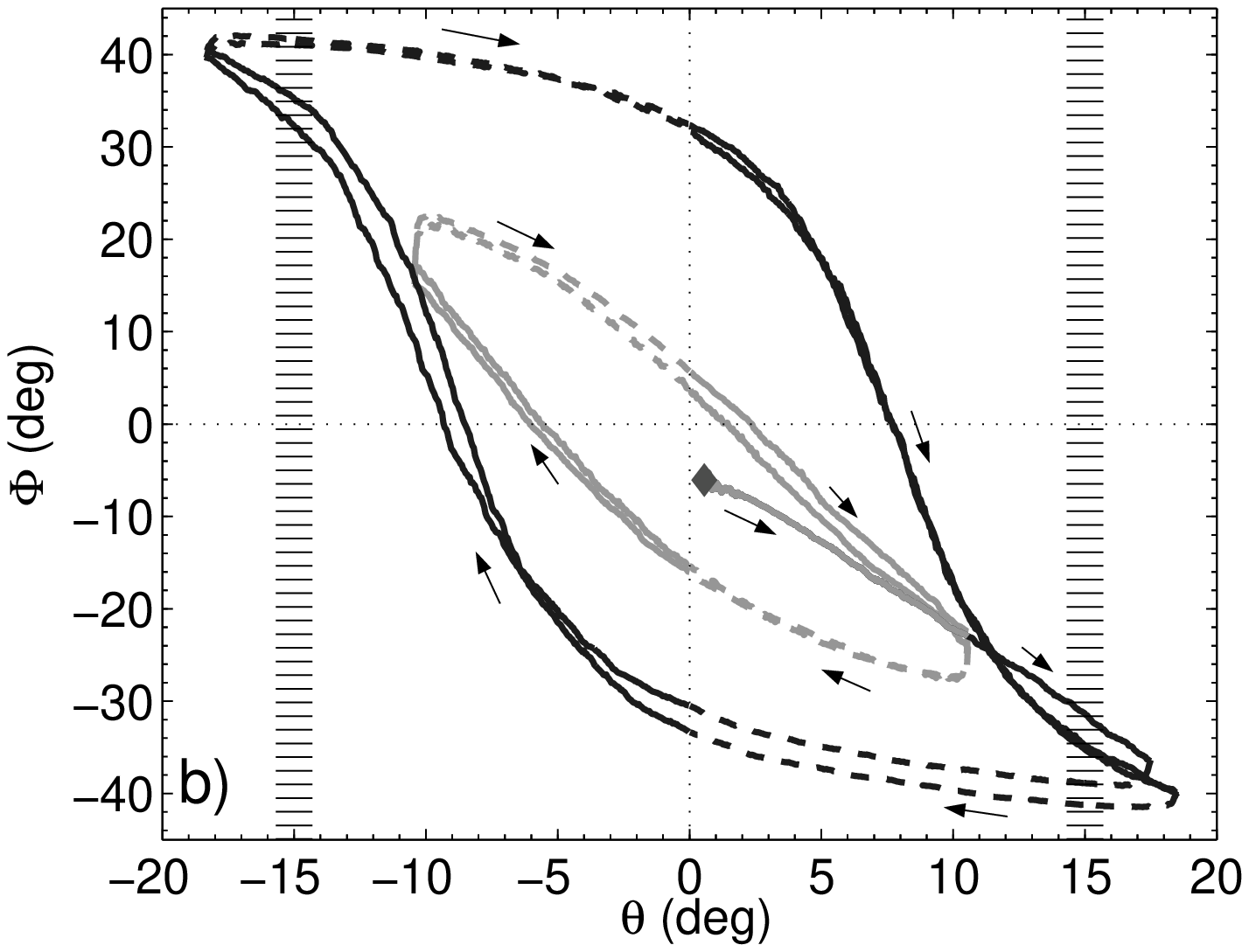,width=\linewidth} }
 \end{minipage}  
 \caption[]{Anisotropic intensity $\Lambda$ ($a$) and fabric anisotropic direction $\Phi$ of 
	the contact network ($b$)
	as a function of the pile slope during the successive cyclic rotations 
	for $\theta_{rev}=10^{\circ}$ and $\theta_{rev}=18^{\circ}$ 
	(same symbols as figure~\ref{stress}). }
 \label{fabric}
\end{figure}
%%fin Fig%%%%

To further analyse the evolution of the micro-structure of the  pile, 
we now consider the statistics of the orientation of contacts. 
To do so, the fabric tensor $\boldsymbol{t}$ is computed using the following definition 
as defined in \cite{Satake1} :
\begin{equation}
t_{ij}= \frac{1}{n^c}\sum_{\alpha =1}^{n^c} n^\alpha_i n^\alpha_j,
\label{eqfabric}
\end{equation}
where $i$ and $j$ denote the space dimensions and $\alpha$ the contact.  
$\vec{n}$ is the unity vector normal to the contact surface, 
and $n^c$ is the total number of contacts. 
The fabric allows to investigate the evolution of the geometry of the contact network, especially 
if contacts are created, opened or modified 
such that their alignment tends towards a privileged direction.
In particular, the anisotropic intensity $\Lambda$ is calculated as :
\begin{equation}
\Lambda=2\times(t_1-t_2),
\end{equation}
where $t_1$ and $t_2$ are respectively the larger and smaller eigenvalues of $\boldsymbol{t}$.
In the case of an isotropic contact network, \eg spatially periodical, the anisotropic intensity $\Lambda=0$.
The direction of the eigenvector of $\boldsymbol{t}$  corresponding to the largest eigenvalue is
the fabric anisotropic direction, whose angle with the direction $\vec{n_S}$  is $\Phi$. 

Figure~\ref{fabric}$a$ displays the evolution of the anisotropic intensity $\Lambda$ 
as a function of the pile slope during the cyclic solicitations 
for $\theta_{rev}=10^{\circ}$ and $\theta_{rev}=18^{\circ}$. 
During small cycles, a rather isotropic state is observed, 
and only small variations of $\Lambda$ with $\theta$ are observed ($\Lambda\simeq0.05$) : 
the geometry of the contact network is only weakly affected by the solicitation 
and remains close to the initial one. 
On the contrary, for large cycles, the anisotropic intensity evolves significantly
showing variations up to the maximum value $\Lambda\simeq0.12$. 
Approaching the unjamming transition obviously strongly modifies the geometry of the contact network, 
\eg by introducing structural anisotropy. 

Figure~\ref{fabric}$b$ displays the evolution of the fabric anisotropic direction $\Phi$ 
as a function of the pile slope during the cyclic solicitations 
for $\theta_{rev}=10^{\circ}$ and $\theta_{rev}=18^{\circ}$.
Irrespective of the value of $\theta_{rev}$, $\Phi$ evolves during the solicitations and 
rotates in the sense opposite to the driving rotation : 
$\Phi$ decreases (resp. increases) when $\theta$ increases (resp. decreases). 
Following this evolution, the fabric anisotropic direction tends to approach that of the free surface and 
for large inclinations, $\Phi$ tends towards the principal stress direction $\Psi=\pm45^{\circ}$. 
This behaviour corresponds to the mechanisms of creation of contacts in the direction of compressive stress, 
and the loss of contacts in the direction of extension. 

Besides these general observations, the contact network evolution exhibits more complex features. 
Both for small and large cycles, 
the fabric exhibits a strongly marked hysteresis, 
probing the irreversible modifications of the disordered micro-structure.
The hysteresis loops of $\Lambda$ and $\Phi$ are very smooth.
Furthermore when the granular pile has approached  the unjamming transition, 
namely for large cycles ($\theta_{rev}=18^{\circ}$), the shape of the hysteresis is dramatically changed, 
and the loop amplified.
This points out the peculiar effect of coherent shearing occurring prior to the unjamming transition 
on the granular packing evolution. 
More particularly, the fabric evolves less rapidly during unloading than during loading 
for large cycles ($\theta_{rev}=18^{\circ}$). 
In other words, the packing seems to be less able to reorganize its structure 
during unloading than during loading when 
having approached  the unjamming transition in its history. 
The same remark applies both to the evolution of the anisotropic intensity $\Lambda$ 
and to the fabric anisotropic direction $\Phi$.
This may be rather unexpected given that the efficiency of rearrangements are 
increased in the case of large cycles, and this during the complete solicitation. 

Note that for large amplitudes ($\theta_{rev}=18^{\circ}$), 
the hysteretic cycles of $\Lambda$ and $\Phi$ are symmetric, 
by contrast with small amplitudes ($\theta_{rev}=10^{\circ}$), for which an asymmetry related to 
the initial fabric anisotropy remains.
Exploring the range of slopes $[\theta_r;\theta_a]$ 
when approaching the unjamming transition allows a profound modification of the contact network 
according to its initial state, by increasing structural disorder. 

Altogether the fabric of the pile (see $Z$, $\Lambda$ and $\Phi$) at an angle $\theta$,   
strongly depends on the solicitation path, as observed for the pile deformation ($\epsilon_V$ and $E_k$).

%----------------------------------------------------------------------
\section{Respective role of strong and weak contacts}  
\label{bimodal}

Dense quasi-static granular media are known to exhibit two complementary contact networks, 
strong and weak, depending on the intensity of the normal force transmitted at contacts \cite{Radjaietal96,Radjaietal98,Aharonov02,Antony00,Staronetal04}. 
Very recently,  the role of the critical contacts in the destabilization of a pile \cite{Staronetal02,StaronRadjai05}, 
as well as in the  elasto-plastic response of a granular packing \cite{Alonsoetal05} has also been underlined.
Still, it was shown that during the slow tilting under gravity of a granular pile, 
critical contacts and more generally high frictional contacts are mainly weak contacts \cite{Staronetal02,StaronRadjai05}. 
Here, in the line of these studies \cite{Radjaietal96,Radjaietal98,Aharonov02,Antony00,Staronetal04} and   as a first step towards the understanding of the role of the 
micro-structure and the heterogeneous nature of contacts in the observed hysteresis, we choose to concentrate on 
the respective contributions of the strong and weak contacts to the hysteretic behaviour of the pile.  
Further works should definitely examine in more details the role of highly frictional contacts in 
the spirit of \cite{StaronRadjai05,Alonsoetal05}. 

Accordingly, the computation of the stress and fabric tensors are now restricted to strong 
or to weak contacts using respectively equation \ref{eqstress} or \ref{eqfabric}. 
In the present work, since a vertical gradient of the contact forces is related to the gravity field, 
a contact at a depth $y$ is defined as strong (resp. weak) 
if it transmits a normal force larger (resp. smaller) than the averaged normal forces at depth $y$.

\subsection{Strong contact network}
%%      Fig 12    %%%%%%%
\begin{figure}[!ht]
 \begin{minipage}{\linewidth}
  \centerline{ \epsfig{file=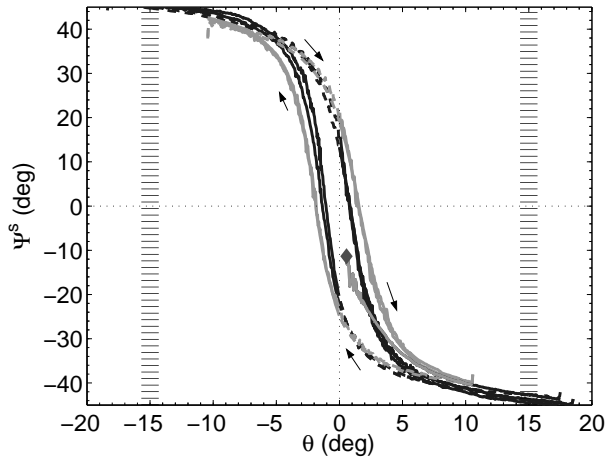,width=\linewidth} }
 \end{minipage}
 \caption[]{Principal stress direction in the strong contact network $\Psi^s$ as a function of 
	the pile slope during the successive loading-unloading cycles for 
	small and large amplitudes (same symbols as figure~\ref{stress}). }
  \label{psiconts}
\end{figure}
%%fin Fig%%%
%%     Fig 13      %%%%%%
\begin{figure}[!ht]
 \begin{minipage}{\linewidth}
  \centerline{ \epsfig{file=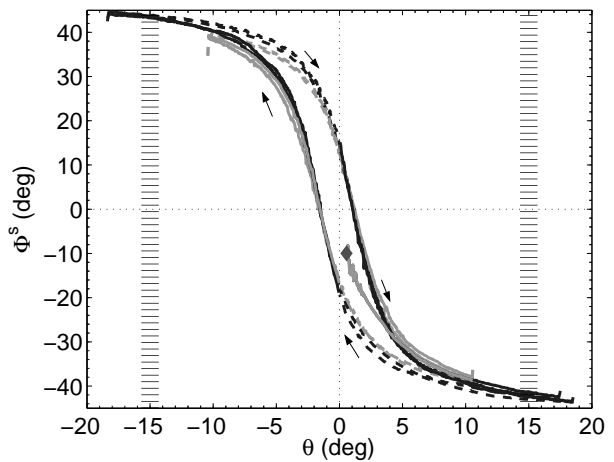,width=\linewidth} }
 \end{minipage}  
 \caption[]{Fabric anisotropic direction of the strong contact network $\Phi^s$ as a function of 
	the pile slope during successive loading-unloading cycles for
	small and large amplitudes (same symbols as figure~\ref{stress}). }
 \label{fabrics}
\end{figure}
%%fin Fig%%%

Figure~\ref{psiconts} displays the principal stress direction  
in the strong contact network $\Psi^s$ as a function of 
the surface slope during small ($\theta_{rev}=10^{\circ}$) and large ($\theta_{rev}=18^{\circ}$) cycles. 
The evolution of $\Psi^s$ is weakly hysteretic as a function of the pile slope and 
the hysteresis is approximately the same irrespective of the amplitude of the solicitation, 
as observed previously for the total contact network (Fig.~\ref{stress}$b$). 
The strong contact network represents by far the largest contribution to the global stress
 \cite{Radjaietal98}. 
Accordingly $\Psi^s$ and $\Psi$ exhibit very similar behaviours. 
 
The fabric anisotropic direction of the strong contact network $\Phi^s$ is displayed as a function 
of the pile slope during the cyclic solicitations in Figure~\ref{fabrics}.
The evolution of $\Phi^s$ exhibits hysteresis during the cycles of both amplitudes. 
However, unlike the previous observations on the fabric of the total contact network (Fig.~\ref{fabric}$b$), 
the size of the hysteresis loops is small 
for $\theta_{rev}=10^{\circ}$ and $18^{\circ}$.
The evolution of $\Phi^s$ is very much correlated with the evolution of $\Psi^s$, and 
the shape of the hysteresis loops of both $\Psi^s$ and $\Phi^s$ are smooth.

During cyclic solicitations, neither the stress nor the fabric of the strong contact network 
is sensitive to the amplitude of the cycles. 

\subsection{Weak contact network}
%%      Fig 14    %%%%%%%
\begin{figure}[!ht]
 \begin{minipage}{\linewidth}
  \centerline{ \epsfig{file=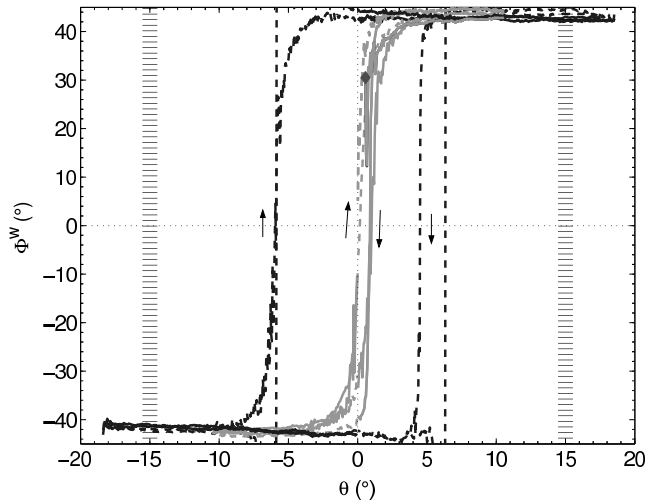,width=\linewidth} }
 \end{minipage} 
  \caption[]{Principal stress direction in the weak contact network $\Psi^w$ as a function
	of the pile slope during the cycles for small and large amplitudes 
	(same symbols as figure~\ref{stress}). }
  \label{psicontw}
\end{figure}
%%fin Fig%%%
%%     Fig 15      %%%%%%
\begin{figure}[!ht]
  \begin{minipage}{\linewidth}
    \centerline{ \epsfig{file=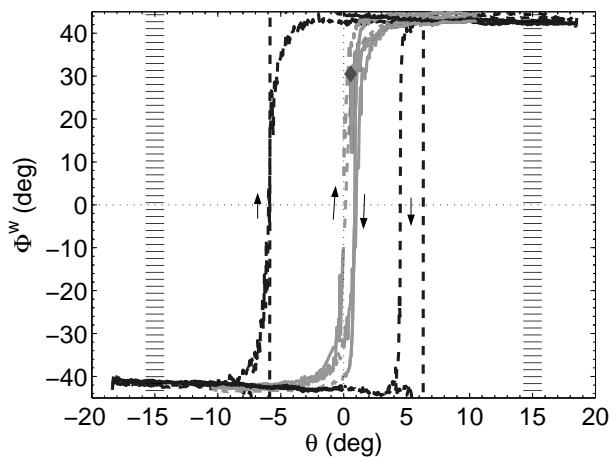,width=\linewidth} }
  \end{minipage} 
\caption[]{Fabric anisotropic direction of the weak contact network $\Phi^w$ as a function of 
	the pile slope $\theta$ during small and large cycles 
	(same symbols as figure~\ref{stress}). }
\label{fabricw}
\end{figure}
%%fin Fig%%%

Figure~\ref{psicontw} displays the principal stress direction
in the weak contact network $\Psi^w$ as a function of 
the surface slope during small ($\theta_{rev}=10^{\circ}$) and large ($\theta_{rev}=18^{\circ}$) cycles. 
The evolution of $\Psi^w$ is strongly influenced by the amplitude of the solicitations.
This qualitative change is not reflected by the total contact network response (Fig.~\ref{stress}$b$) 
due to the dominant contribution of the strong contact network to the global stress \cite{Radjaietal98}. 

Figure~\ref{fabricw} displays the evolution of
the fabric anisotropic direction of the weak contact network $\Phi^w$ 
as a function of the pile slope for small 
($\theta_{rev} =10^{\circ}$) and large ($\theta_{rev} =18^{\circ}$) cycles.
The response of the weak contact network is again strongly dependent 
on the cycles amplitude $\theta_{rev}$, 
by contrast with the response of the strong contact network (Fig.~\ref{fabrics}). 
For small cycles ($\theta_{rev}=10^{\circ}$), $\Phi^w$ exhibits nearly no hysteresis, and 
remains approximately always normal to $\Phi^s$, as observed in the case
of continuous loading \cite{StaronRadjai05}.
On the contrary, for large cycles ($\theta_{rev}=18^{\circ}$), 
a strong hysteretic behaviour is observed,  which consists of a premature 
rotation of $\Phi^w$ upon reversal : $\Phi^w$ switches beforehand during unloading in comparison with loading. 
This effect leads to a remarkable orientation of the fabric of the two contact networks :
$\Phi^w$ and $\Phi^s$ are equal during a part of the cycle, 
attesting the temporary alignment of the privileged direction of the two contact networks. 
The particularly early evolution of the weak contact network geometry  
when the pile has approached the unjamming transition,   
is actually responsible for the delay observed in the reorganization of the total contact network.
As a consequence, upon reversal, the packing keeps the memory of the previous  
orientation of the total contact network during a large part of the cycle (Fig.~\ref{fabric}$b$).

Contrary to the smooth shape of the hysteresis loops in the strong contact network (Fig.~\ref{fabrics})
and in the total contact network (Fig.~\ref{fabric}$b$),
the hysteresis loops in the weak contact network (Fig.~\ref{fabricw}) exhibit an abrupt macroscopic jump.
It should also be mentioned that the weak contact network response exhibits large fluctuations 
from one realization to another one. 
These fluctuations are strongly enhanced close to the unjamming transition.
Evidences of these fluctuations are for instance the noisy signal for $\Psi^w$ (Fig.~\ref{psicontw}) 
and the absence of superposition of the hysteretic loops for $\Phi^w$ (Fig.~\ref{fabricw}).

Altogether, in the weak contact network by contrast with the strong one, both stress and fabric 
are very sensitive to the amplitude of the cyclic solicitations. 
Two conclusions can be drawn from these observations : 
\begin{itemize}
 \item In the limit of small solicitations, \eg for cycles with $\theta_{rev}=10^{\circ}$, 
 the hysteretic behaviour 
 of the granular pile when looking at both stress and fabric state  
 is dominated by the strong contact network contribution ; 
 \item For large cycles, with $\theta_{rev}>\theta_r$, namely when the pile approaches the unjamming transition, 
 the response of the weak contact network becomes strongly affected, 
 and the hysteretic behaviour of the pile is strongly
 modified despite the lack of modification of the response of the strong contact network. 
\end{itemize}
Accordingly the hysteretic behaviour of the pile is the result of a complex interplay between
the strong and weak contact networks, whose relative contributions strongly depend on 
the amplitude of the cycles.

%----------------------------------------------------------------------
\section{Discussion and conclusion}
\label{discussion}

The successive tiltings of the granular pile from different initial conditions 
up towards the avalanche angle allow to demonstrate that 
the density and the correlation length of the critical contacts 
are relevant dynamic response functions to describe the metastable evolution of the packing  
towards the unjamming transition under external loading. 
 Due to the transient nature of the critical contact state, these quantities are representative of actual loading  of the pile, but they do not characterize its plastic state.

The response of the granular packing to the quasi-static cyclic solicitations reveals that the evolution
of the critical contacts is hysteretic when the pile has approached the unjamming transition. 
At the transition from loading to unloading, 
 critical contacts can not any more sustain the sliding condition, so that 
the frictional state of the packing is immediately removed,
attesting the end of coherent shearing, 
as observed in the case of experimental  shear reversal \cite{Toiyaetal04,UttBeh04}.

When looking for such memory effects on other features of the pile, namely on   
stress, strain and fabric, they exhibit different responses. 
The stress state exhibits weak hysteretic evolutions as a function of the slope, 
and the hysteresis reveals to be independent of the amplitude of the cycles. 
On the contrary, the strain and the fabric evolutions of the pile, related to structural disorder, 
are very sensitive to the amplitude of the cycles. 
These observations will hardly be explained  by a simple constitutive law 
 relating only stress and strain of the granular pile :
such a relation should be not so trivial, but should take into account the anisotropy of the pile micro-structure \cite{ThorntonBarnes86,BathurstRothenburg88,Vardoulakis89,Nemat00,Madadietal04,Alonsoetal05}. 
The observation of piles with the same stress state but characterized by different  granular fabrics, 
would suggest that stress induced by gravity are transmitted in the pile whatever the geometry of the total contact network, and whatever the  structural disorder  level.  

Further analyses of the strong and weak contacts contributions to the macroscopic behaviour
allow to clarify these intriguing observations.
In the strong contact network, neither the stress nor the fabric depends on the solicitation amplitude.
In the weak contact network, both the stress and the fabric are sensitive to the cyclic amplitude.
The apparent mismatch between stress and fabric at the scale of the total contact network does not survive
when separating weak and strong contact networks contributions.
It simply comes from the strong contact network weight in the overall stress \cite{Radjaietal98}.
Accordingly, the above analyses stress the relevance of a constitutive law which specifies the two-phases 
 nature --weak and strong-- of granular media.
In the strong contact network, stress and fabric are very much synchronised,  and a rather simple relation  
could relate them. 
The behaviour of the weak contact network is much more complex. 
A strong memory of the unjamming transition affects the response  
of the weak contact network during the full solicitation, 
probably due to high sensibility of weak contacts to coherent shearing 
occurring prior to the unjamming transition.
Finally the observed fluctuations in the weak contact network are presumably related to long-range correlations. 
These correlations within the pile allow a local rearrangement of grains 
to further propagate and to affect essentially the weak contacts, 
due to their highest frictional mobilization, as recently shown in \cite{Staronetal02,StaronRadjai05}. 
This high sensibility of the weak contacts to local perturbations enables them to evolve collectively and 
to change the micro-structure of the packing.
A signature of the localization of the correlations in the weak contact network  
could be seen in the particular shape of the hysteresis loops. 
Whereas the loops are very smooth in the total and strong contact networks, 
the hysteresis loops in the weak contact network exhibit abrupt macroscopic jumps.
These respective delay and instantaneous responses could be related to the relative contribution of structural disorder and 
finite-size interactions or correlations, as suggested in \cite{DahmenSethna96,Jiangetal99}.
In this picture, the hysteresis with a jump in the weak contact network would suggest that the size of correlations
is larger than the typical length-scale of the structural disorder in  this phase, 
contrary to the predominant role of disorder in the micro-structure of the packing. 
 Altogether, such complex behaviours --spatial correlations and memory effects-- call for more 
elaborated constitutive laws as in  \cite{ThorntonBarnes86,BathurstRothenburg88,Vardoulakis89,Nemat00,Madadietal04,Alonsoetal05}. 

To conclude, the behaviour of the pile is strongly modified close to the unjamming transition, 
as experimentally observed \cite{DaerrDouady99,Deboeufetal03,Kablaetal04},  
coinciding with important irreversible modifications of the micro-structure.
These modifications result from a specific solicitation of weak contacts : 
these latter play a considerable role in the behaviour of the pile 
despite their marginal contribution to the stress state. 
The present paper demonstrates the relevance of a two-phases description, 
not only for the destabilization process, but also for the quasi-static rheology, as soon as
the granular sample has approached the unjamming transition in its past history.
Further studies should investigate the complex behaviour of the weak phase 
in order to extract a constitutive relation  for granular materials,  accounting for their two-phases nature. 

\vspace{1cm}

{\bf Acknowledgments :}  We wish to thank the French Groupe de Recherche Milieux Divis{\'e}s, 
and more especially J. H. Snoeijer, P. Claudin, J.-N. Roux and E. Kolb  for their discussions on the present work. 
L. Staron acknowledges  the Marie Curie European Grant $500511$ for financial support. 

%----------------------------------------------------------------------
\bibliographystyle{unsrt}
\bibliography{biblio}

\begin{thebibliography}{10}

\bibitem{Sollichetal97}
P.~Sollich, F.~Lequeux, P.~Hebraud, and M.~E. Cates.
\newblock Rheology of soft glassy materials.
\newblock {\em Phys. Rev. Lett.}, 78(10), 1997.

\bibitem{Roux00}
J.-N. Roux.
\newblock Geometric origin of mechanical properties of granular materials.
\newblock {\em Phys. Rev. E}, 61(6802), 2000.

\bibitem{Troadecetal02}
H.~Troadec, F.~Radjai, S.~Roux, and J.~C. Charmet.
\newblock Model for granular texture with steric exclusion.
\newblock {\em Phys. Rev. E}, 66(41305), 2002.

\bibitem{JaeNagBeh96}
H.~M. Jaeger, S.~R. Nagel, and R.~P. Behringer.
\newblock Granular solids, liquids, and gases.
\newblock {\em Rev. Mod. Phys.}, 68(4), 1996.

\bibitem{OdaIwashita99}
M.~Oda and K.~Iwashita.
\newblock {\em Mechanics of Granular Materials: an Introduction}.
\newblock A.A. Balkema, Rotterdam, 1999.

\bibitem{deGennes99a}
P.G. de~Gennes.
\newblock Granular matter: a tentative view.
\newblock {\em Rev. Mod. Phys.}, 71(2), 1999.

\bibitem{Rajchenbach00}
J.~Rajchenbach.
\newblock Granular flows.
\newblock {\em Advances in Physics}, 49, 2000.

\bibitem{NematHori99}
S.~Nemat-Nasser and M.~Hori.
\newblock {\em Micromechanics: overall properties of heterogeneous solids}.
\newblock Elsevier, Amsterdam, 2nd edition, 1999.

\bibitem{RadjaiRoux04}
F.~Radjai and S.~Roux.
\newblock Contact dynamics study of 2{D} granular media: Critical states and
  relevant internal variables.
\newblock In H.~Hinrichsen and D.E. Wolf, editors, {\em The physics of granular
  media}. Wiley, Berlin, 2004.

\bibitem{Knightetal95}
J.B. Knight, C.G. Fandrich, C.N. Lau, H.M. Jaeger, and S.~Nagel.
\newblock Density relaxation in a vibrated granular material.
\newblock {\em Phys. Rev. E}, 3957(51), 1995.

\bibitem{OHernetal01}
C.~S. O'Hern, S.~A. Langer, J.~A. Liu, and A.~J. Nagel.
\newblock Force distributions near jamming and glass transitions.
\newblock {\em Phys. Rev. Lett.}, 86(111), 2001.

\bibitem{HartleyBehringer03}
R.~R. Hartley and R.~P. Behringer.
\newblock Logarithmic rate dependence of force networks in sheared granular
  materials.
\newblock {\em Nature}, 421:928, 2003.

\bibitem{GdrMidi04}
GDR MiDi.
\newblock On dense granular flows.
\newblock {\em Eur. Phys. J. E}, 14:341, 2004.

\bibitem{LiuNagel98}
A.~J. Liu and S.~R. Nagel.
\newblock Jamming is not just cool anymore.
\newblock {\em Nature}, 396:21, 1998.

\bibitem{OHernetal03a}
C.~S. O'Hern, L.~E. Silbert, A.~J. Liu, and S.~R. Nagel.
\newblock Jamming at zero temperature and zero applied stress: The epitome of
  disorder.
\newblock {\em Phys. Rev. E}, 68(11306), 2003.

\bibitem{Berthieretal01}
L.~Berthier, L.~F. Cugliandolo, and J.~L. Iguain.
\newblock Glassy systems under time-dependent driving forces: Application to
  slow granular rheology.
\newblock {\em Phys. Rev. E}, 63(051302), 2001.

\bibitem{Pouetal03}
O.~Pouliquen, M.~Belzons, and M.~Nicolas.
\newblock Fluctuating particle motion during shear induced compaction.
\newblock {\em Phys. Rev. Lett.}, 91(014301), 2003.

\bibitem{Conetal04}
A.~Coniglio, A.~Fierro, H.J. Herrmann, and M.~Nicodemi.
\newblock {\em Unifying concepts in granular media and glasses}.
\newblock Elsevier, Amsterdam, 2004.

\bibitem{MartyDauchot05}
G.~Marty and O.~Dauchot.
\newblock Subdiffusion and cage effect in a sheared granular material.
\newblock {\em Phys. Rev. Lett.}, 94(015701), 2005.

\bibitem{Edwards94}
S.~F. Edwards.
\newblock {\em Granular Matter : An interdisciplinary Approach}.
\newblock Springer, New-York, a. mehta edition, 1994.

\bibitem{Onoetal02}
I.~K. Ono, C.~S. O'Hern, D.J. Durian, S.~A. Langer, A.~J. Liu, and S.~R. Nagel.
\newblock Effective temperatures of a driven system near jamming.
\newblock {\em Phys. Rev. Lett.}, 89(095703), 2002.

\bibitem{Deboeufetal03}
S.~Deboeuf, E.~M. Bertin, E.~Lajeunesse, and O.~Dauchot.
\newblock Jamming transition of a granular pile below the angle of repose.
\newblock {\em Eur. Phys. J. B}, 36:105, 2003.

\bibitem{UttBeh04}
B.~Utter and R.P. Behringer.
\newblock Transients in sheared granular matter.
\newblock {\em Eur. Phys. J. E}, 14:373, 2004.

\bibitem{MehtaBarker01}
A.~Mehta and G.C. Barker.
\newblock Bistability and hysteresis in tilted piles.
\newblock {\em Europhys. Lett.}, 56(5), 2001.

\bibitem{Metetal02}
G.~Metcalfe, S.G.K. Tennakoon, L.~Kondic, D.G. Shaeffer, and R.P. Behringer.
\newblock Granular friction, {C}oulomb failure, and fluid-solid transition for
  horizontally shaken granular materials.
\newblock {\em Phys. Rev. E}, 65(031302), 2002.

\bibitem{DanBeh04}
K.~E. Daniels and R.~P. Behringer.
\newblock Hysteresis and competition between disorder and crystallization in
  sheared and vibrated granular flow.
\newblock {\em Phys. Rev. Lett.}, 94(168001), 2005.

\bibitem{Staronetal02}
L.~Staron, J.-P. Vilotte, and F.~Radjai.
\newblock Preavalanche {i}nstabilities in a {g}ranular {p}ile.
\newblock {\em Phys. Rev. Lett.}, 89(204302), 2002.

\bibitem{Kablaetal04}
A.~Kabla, G.~Debregeas, J.-M. di~Meglio, and T.~J. Senden.
\newblock X-ray observation of micro-failures in granular piles approaching an
  avalanche.
\newblock {\em Europhys. Lett.}, 10165(4), 2005.

\bibitem{Siletal05}
L.E. Silbert, A.J. Liu, and S.R. Nagel.
\newblock Vibrations and diverging length scales near the unjamming transition.
\newblock {\em Phys. Rev. Lett.}, 95(098301), 2005.

\bibitem{Jaegeretal89}
H.~M. Jaeger, C.~Liu, and S.~R. Nagel.
\newblock Relaxation at the {A}ngle of {R}epose.
\newblock {\em Phys. Rev. Lett.}, 62(40), 1989.

\bibitem{DaerrDouady99}
A.~Daerr and S.~Douady.
\newblock Two types of avalanche behaviour in granular media.
\newblock {\em Nature}, 399:241, 1999.

\bibitem{Staronetal04}
L.~Staron, J.-P. Vilotte, and F.~Radjai.
\newblock Multiscale {a}nalysis of the {s}tress {s}tate in a {g}ranular {s}lope
  in {t}ransition to {f}ailure.
\newblock {\em cond-mat/0409619}, 2004.

\bibitem{StaronRadjai05}
L.~Staron and F.~Radjai.
\newblock Friction vs {T}exture at the {A}pproach of a {G}ranular {A}valanche.
\newblock {\em cond-mat/0505097}, 2005.

\bibitem{Radjaietal98}
F.~Radjai, D.~E. Wolf, M.~Jean, and J.-J. Moreau.
\newblock Bimodal character of stress transmission in granular packings.
\newblock {\em Phys. Rev. Lett.}, 80:61, 1998.

\bibitem{Radjaietal96}
F.~Radjai, M.~Jean, J.J. Moreau, and S.~Roux.
\newblock Force distribution in dense two-dimensional granular systems.
\newblock {\em Phys. Rev. Lett.}, 77:274, 1996.

\bibitem{Nasunoetal97}
S.~Nasuno, A.~Kudrolli, and J.P. Gollub.
\newblock Friction in granular layers: Hysteresis and precursors.
\newblock {\em Phys. Rev. Lett.}, 79(949), 1997.

\bibitem{Toiyaetal04}
M.~Toiya, J.~Stambaugh, and W.~Losert.
\newblock Transient and oscillatory granular shear flow.
\newblock {\em Phys. Rev. Lett.}, 93(88001), 2004.

\bibitem{Mueggenburg04}
N.~W. Mueggenburg.
\newblock Behavior of granular materials under cyclic shear.
\newblock {\em Phys. Rev. E}, 71(031301), 2005.

\bibitem{AlonsoHerrmann04}
F.~Alonso-Marroqu\'in and H.J. Herrmann.
\newblock Ratcheting of {g}ranular {m}aterials.
\newblock {\em Phys. Rev. Lett.}, 92(054301), 2004.

\bibitem{Garcia05}
R.~Garc\'ia-Rojo, F.~Alonso-Marroqu\'in, and H.~J. Herrmann.
\newblock Characterization of the material response in the granular ratcheting.
\newblock {\em cond-mat/0505507}, 2005.
\newblock accepted in Phys. Rev. E.

\bibitem{JeanandMoreau92}
M.~Jean and J.-J. Moreau.
\newblock In {\em Proceedings of Contact Mechanics, International Symposium}.
  PPUR, 1992.

\bibitem{KruytRothenburg96}
N.P. Kruyt and L.~Rothenburg.
\newblock Micromechanical definition of the strain tensor for granular
  material.
\newblock {\em J. App. Mech}, 706(118), 1996.

\bibitem{Porionetal93}
P.~Evesque, D.~Fargeix, P.~Habib, M.P. Luong, and P.~Porion.
\newblock Pile density is a control parameter of sand avalanches.
\newblock {\em Phys. Rev. E}, 47:2326, 1993.

\bibitem{Nicolasetal00}
M.~Nicolas, P.~Duru, and O.~Pouliquen.
\newblock Compaction of a granular material under cyclic shear.
\newblock {\em Eur. Phys. J. E}, 3:309, 2000.

\bibitem{Joer98}
H.A. Joer, J.~Lanier, and M.~Fahey.
\newblock Deformation of granular materials due to rotation of principal axes.
\newblock {\em Geotechnique}, 48(5), 1998.

\bibitem{Alonsoetal05}
F.~Alonso-Marroqu\'in, S.~Luding, H.J. Herrmann, and I.~Vardoulakis.
\newblock Role of anisotropy in the elastoplastic response of a polygonal
  packing.
\newblock {\em Phys. Rev. E}, 71(051304), 2005.

\bibitem{Madadietal04}
M.~Madadi, O.~Tsoungui, M.~Lätzel, and S.~Luding.
\newblock {\em Int. J. Solids Struct.}, 41(2563), 2004.

\bibitem{ThorntonBarnes86}
C.~Thornton and D.~J. Barnes.
\newblock {\em Acta Mech.}, 64(45), 1986.

\bibitem{BathurstRothenburg88}
R.~J. Bathurst and L.~Rothenburg.
\newblock {\em J. Appl. Mech.}, 55, 1988.

\bibitem{Nemat00}
S.~Nemat-Nasser.
\newblock {\em J. Mech. Phys. Solids}, 28, 2000.

\bibitem{Siletal02}
L.~E. Silbert, D.~Ertas, G.~S. Grest, T.~C. Halsey, and D.~Levine.
\newblock Geometry of frictionless and frictional sphere packings.
\newblock {\em Phys. Rev. E}, 65(031304), 2002.

\bibitem{Satake1}
M.~Satake.
\newblock In {\em Deformation and Failure of Granular Materials}. A.A. Balkema,
  1982.

\bibitem{Antony00}
S.~J. Antony.
\newblock Evolution of force distribution in three-dimensional granular media.
\newblock {\em Phys. Rev. E}, 63(011302), 2000.

\bibitem{Aharonov02}
E.~Aharonov and D.~Sparks.
\newblock Shear profiles and localization in simulations of granular materials.
\newblock {\em Phys. Rev. E}, 65(051302), 2002.

\bibitem{Vardoulakis89}
I.~Vardoulakis.
\newblock {\em Ing.-Arch.}, 59(106), 1989.

\bibitem{DahmenSethna96}
K.~Dahmen and J.~P. Sethna.
\newblock Hysteresis, avalanches, and disorder-induced critical scaling: A
  renormalization-group approach.
\newblock {\em Phys. Rev. B}, 53(14872), 1996.

\bibitem{Jiangetal99}
Y.~Jiang, P.~J. Swart, A.~Saxena, M.~Asipauskas, and J.~A. Glazier.
\newblock Hysteresis and avalanches in two-dimensional foam rheology
  simulations.
\newblock {\em Phys. Rev. E}, 59(5819), 1999.

\end{thebibliography}

\end{document}